\documentstyle[12pt]{article}
\begin{document}
\thispagestyle{empty}

\begin{center}
\LARGE \tt \bf {Gravitational stability of inflaton and torsion in Einstein-Cartan-Klein-Gordon cosmology}
\end{center}

\vspace{2.5cm}

\begin{center} {\large 

L.C. Garcia de Andrade\footnote{Departamento de
F\'{\i}sica Te\'{o}rica - Instituto de F\'{\i}sica - UERJ

Rua S\~{a}o Fco. Xavier 524, Rio de Janeiro, RJ

Maracan\~{a}, CEP:20550-003 , Brasil.E-mail:garcia@dft.if.uerj.br}}
\end{center}
\vspace{2cm}

\begin{abstract}
Gravitational stability of torsion and inflaton potential in a four-dimensional spacetime de Sitter solution in scalar-tensor cosmology where Cartan torsion propagates is investigated in detail. Inflaton and torsion evolution equations are derived by making use of a Lagrangean method. Stable and unstable modes for torsion and inflatons are found. Present astrophysical observations favour a stable mode for torsion since this would explain why no relic torsion imprint has been found on the Cosmic Background Radiation in the universe. 
\end{abstract}

\newpage
\section{Introduction}
\indent
 Some attempts have been made recently \cite{1,2,3,4} to investigate the cosmological perturbations  \cite{5} in spacetimes with torsion. The simplest example is the C.Stornaiolo et al.  \cite{1} which considered gauge invariant approach by simply considered the extension of general relativistic small perturbations and substituting into it the effective matter density and pressure into the evolution equations where the effective quantities bring the spin-torsion density on the expressions. Earlier Piskareva and Obukhov \cite{2} have considered the vector (rotational), scalar (matter density) and tensorial (gravitational waves) small perturbations of spinning fluid matter without nevertheless without torsion. More recently Garcia de Andrade \cite{3} has considered the stability of inflatons torsion and metric fluctuations around de Sitter solutions in Einstein-Cartan-Klein-Gordon (ECKG) cosmology. In this work we make use of the technique of building the evolution equations from the small perturbations directly and no attempt of using the traditional Lagrangean method was made. In this Letter we follow the approach considered by Maroto and Shapiro \cite{4} who investigated the existence and stability of de Sitter inflationary solutions for the string-inspired fourth-derivative gravity theories with torsion and dilatons in arbitrary $D-dimensional spacetime$. One of the main features of their work as far as the study of the role of torsion is concerned is that Cartan torsion field is nondynamic in the sense that torsion obeys an algebraic equation and not a differential equation. On the other hand in their theory de Sitter solution fluctuates to a more general solution of the string action. In this Letter we propose to work with the Lagrangean method to investigate a scalar-tensor action where now the Lagrangean would depend also on torsion and not only to the metric cosmic scale factor and inflaton field. The present letter is organized as follows. In the section $2$ we present the derive the field equation of the action of this scalar-tensor cosmology and found a simple solution. In section $3$ we investigate the fluctuations on the inflaton and torsion and derive the evolution equations for both fields. We also show by solving this equation in a simple case that torsion and inflatons possess stable and unstable modes. Our solutions and equations are clearly distinct to the ones obtained previously by Maroto and Shapiro in reference $2$. Section $4$ presents some brief discussion.
Of course during the paper we consider inflatons instead of dilatons since we are dealing with four-dimensional inflationary cosmology. It is also important to note that here we do not attempt to find an exct solution of the ECKG field equations of gravity \cite{5} but only to investigate the behaviour and stability of inflatons and torsion against the background de Sitter solution, this is possible since we have shown very recently \cite{3} that ECKG equation is reduced to Einstein-de Sitter equation around the inflaton constant background field ${\phi}_{0}$.

\section{Inflaton and Torsion Equations in Inflationary Cosmology.}
\indent
In this section we present the torsion and inflaton sections derived from the action \cite{4}

\begin{equation}
S_{M} = \frac{2}{k^{2}}\int{d^{4}x \sqrt{g}[ - R({\Gamma})(1+k{\phi}^{2}) + 4({\partial}{\phi})^{2}]}
\label{1}
\end{equation}
where $ g $ is the determinant of the de Sitter metric and $R({\Gamma})$ is the non-Riemannian Ricci scalar given by

\begin{equation}
R = -6 H^{2}_{0}- \dot{T}- 6 T^{2}(t)
\label{2}
\end{equation}
where ${\phi}$ is the inflaton potential $T(t)$ is the torsion zero component of the torsion vector while $H_{0}$ is the de Sitter expansion factor which reads
\begin{equation}
ds^{2} = dt^{2} - e^{2H_{0}t}(dx^{2}+dy^{2}+dz^{2})
\label{3}
\end{equation}
Therefore the above action reads
\begin{equation}
S_{M} = \frac{2}{k^{2}}\int{d^{4}x  e^{3H_{0}t}[- 6(H^{2}_{0}+\frac{1}{6}\dot{T}+{T}^{2})(1+k{\phi}^{2}) + 4{\dot{\phi}}^{2}]}
\label{4}
\end{equation}
Writing this action in the form 
\begin{equation}
S = \frac{2}{k^{2}}\int{dt e^{3H_{0}t} L(T,\dot{T},{\phi},\dot{\phi})}
\label{5}
\end{equation}

here $L$ represents the Lagrangean function. Variation of this action leads to the following field equations for torsion and inflaton
\begin{equation}
3H_{0}\frac{{\partial}{L}}{{\partial}{\dot{T}}}+ \frac{d}{dt}\frac{{\partial}L}{{\partial}{\dot{T}}}-\frac{{\partial}{L}}{{\partial}{T}}=0
\label{6}
\end{equation}
By the symmetry of the Lagrangean the expression for the inflaton field is given by

\begin{equation}
3H_{0}\frac{{\partial}{L}}{{\partial}{\dot{\phi}}}+ \frac{d}{dt}\frac{{\partial}L}{{\partial}{\dot{\phi}}}-\frac{{\partial}{L}}{{\partial}{\phi}}=0
\label{7}
\end{equation}
Substitution of Lagrangean L in equation (\ref{4}) one obtains 
\begin{equation}
\ddot{\phi}+24H_{0}\dot{\phi}-2kR{\phi}=0
\label{8}
\end{equation}
This equation has a simple solution with the ansatz ${\phi}=e^{{\alpha}t}$. This ansatz reduces the inflaton equation to an algebraic equation 
\begin{equation}
{\alpha}^{2}+24H_{0}{\alpha}-2kR=0
\label{9}
\end{equation}
After some algebra substitution of expression 
\begin{equation}
{\phi} \rightarrow {\phi}_{0}+ {\delta}{\phi}
\label{10}
\end{equation}
into equation for the evolution of inflaton yields
\begin{equation}
{\delta}\ddot{\phi}+ 24H_{0}{\delta}\dot{\phi}-2kR{\delta}{\phi}-2k{\delta}R{\phi}=0
\label{11}
\end{equation}
Before complete this evolution equation we need to compute torsion fluctuation equation. This yields 
\begin{equation}
T(t)= -\frac{1}{24}H_{0}-\frac{1}{12}\frac{d}{dt}ln(1+k{\phi}^{2})
\label{12}
\end{equation}
Let us now considered the fluctuation of the Cartan torsion against a constant background torsion $T_{0}$ given by
\begin{equation}
T \rightarrow T_{0}+ {\delta}T
\label{13}
\end{equation}
Substitution of this expression into the equation (\ref{11}) yields
\begin{equation}
{\delta}T(t)= -\frac{1}{12}\frac{d}{dt}ln(1+k[{{\phi}_{0}}^{2}+2{\phi}_{0}{\delta}{\phi}])=-\frac{1}{12(1+k[2{\phi}_{0}{\delta}{\phi}])}(2{\phi}_{0}\frac{d}{dt}{\delta}{\phi})
\label{14}
\end{equation}
and 
\begin{equation}
T_{0}(t)= -\frac{1}{24}H_{0}
\label{15}
\end{equation}
where we also consider the small perturbation of the inflaton field above. Now substitution of ${\delta}T$ and its first derivative in ${\delta}R$ in expression  (\ref{11}) yields
\begin{equation}
(1-{\gamma}){\delta}\ddot{\phi}+ 24(H_{0}+ k{\phi}_{0}T_{0}){\gamma}{\delta}\dot{\phi}-2kR_{0}{\delta}{\phi}-2kR_{0}=0
\label{16}
\end{equation}
to simplify the solution of this equation we addopt the following constraint approximation $R_{0}=0$ which yields the constraint $\frac{d}{dt}T|_{0} = 6(H_{0}^{2}+T_{0}^{2})$. With this constraint the evolution equation for inflaton small perturbations is easily solved with the ansatz ${\delta}{\phi}= e^{{\lambda}t}$. This yields the following carachteristic equation
\begin{equation}
{\lambda}=-24(H_{0}^{2}+k{\phi}_{0}T_{0}){\gamma}
\label{17}
\end{equation}
Therefore the inflaton field possess a stable mode against small perturbations in de Sitter background metric. Substitution of expression (\ref{17}) into (\ref{14}) we note that torsion is also stable against small perturbations which implies the is not unstable which would explain physically why torsion cannot be found as a cosmic relic in present universe observations \cite{6}. 
\section*{Acknowledgements}
I would like to express my gratitude to Professors Yuri Obukhov and Ilya Shapiro for helpful discussions and sugestions on the subject of this paper. Financial support from CNPq. is gratefully acknowledged.


\newpage
\begin{thebibliography}{5}
\bibitem{1}C. Stornaiolo et al. Il Nuovo Cimento B (1997).
\bibitem{2}Yu Obukhov and Piskareva,JETP (1990).
\bibitem{3}L.C.Garcia de Andrade,On the stability of de Sitter inflationary background in scalar-tensor cosmology,astro-ph/0021167.
\bibitem{4}Maroto and I.Shapiro,Phys. Lett. B (1997).
\bibitem{5}C. Stornaiolo,Sulle Cosmologia nella ECKS,(1985)PhD thesis Naples University.
\end{thebibliography}
\end{document}